\documentclass[12pt]{article}
\usepackage[margin=1.0in]{geometry}
\usepackage{graphicx}
\usepackage{float}
\usepackage{wrapfig}
\usepackage{caption}
\usepackage{amssymb}
\usepackage{amsmath}
\usepackage{natbib}
\usepackage{url}
\usepackage{aas_macros}

\begin{document}
\begin{titlepage}
    \centering
    \hphantom{.1mm}
    \vfill
{\scshape \LARGE{Radio Time-Domain Signatures of Magnetar Birth}\par}
    \vspace{1cm}
{\large{\bfseries A White Paper for the Astro2020 Decadal Review }} \\
    \vfill


    \vfill

	\begin{tabular}{cc}
Casey J.~Law & \textit{UC Berkeley}, \url{claw@astro.berkeley.edu} \\
Ben Margalit & \textit{UC Berkeley}, \url{benmargalit@berkeley.edu} \\
Nipuni T. Palliyaguru & \textit{Arecibo Observatory}, \url{nipunipalliyaguru9@gmail.com} \\
Brian D.~Metzger & \textit{Columbia University}, \url{bdm2129@columbia.edu} \\
Lorenzo Sironi & \textit{Columbia University}, \url{lsironi@astro.columbia.edu} \\
Yong Zheng & \textit{UC Berkeley}, \url{yongzheng@berkeley.edu} \\
Edo Berger & \textit{Harvard University}, \url{eberger@cfa.harvard.edu} \\
Raffaella Margutti & \textit{Northwestern Univ.}, \url{raffaella.margutti@northwestern.edu} \\
Andrei Beloborodov & \textit{Columbia University}, \url{amb2046@columbia.edu} \\
Matt Nicholl & \textit{University of Edinburgh}, \url{mrn@roe.ac.uk} \\
Tarraneh Eftekhari & \textit{Harvard University}, \url{tarraneh.eftekhari@cfa.harvard.edu} \\
Indrek Vurm & \textit{Tartu Observatory}, \url{indrek.vurm@ut.ee} \\
Peter Williams & \textit{Center for Astrophysics, AAS}, \url{pwilliams@cfa.harvard.edu } \\
	\end{tabular}

    \vfill

	\noindent \textbf{Thematic Areas:} \hspace*{20pt} 
	$\square$ Planetary Systems \hspace*{10pt} 
	$\square$ Star and Planet Formation \hspace*{10pt}\linebreak
    $\rlap{$\checkmark$} \square$ Formation and Evolution of Compact Objects \hspace*{10pt} 
    $\rlap{$\checkmark$} \square$ Cosmology and Fundamental Physics \linebreak
    $\square$  Stars and Stellar Evolution \hspace*{10pt} $\square$ Resolved Stellar Populations and their Environments  \linebreak
    \hspace*{30pt} $\rlap{$\checkmark$} \square$    Galaxy Evolution   \hspace*{10pt} 
    $\rlap{$\checkmark$} \square$    Multi-Messenger Astronomy and Astrophysics \hspace*{65pt} \linebreak
	\vfill
    \end{titlepage}

\begin{wrapfigure}[23]{r}{0.4\textwidth}
  \centering
  \vspace{-15pt}
  \includegraphics[bb=0in 1in 8.5in 9.5in, width=0.4\columnwidth]{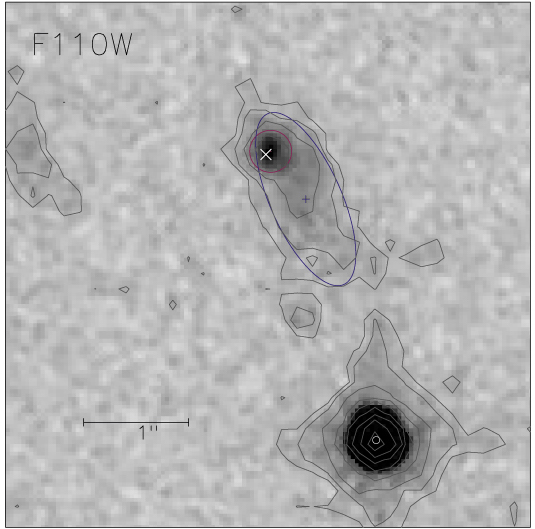}
  \caption{HST observation of the FRB 121102 host galaxy at $z=0.1927$ \citep{2017Natur.541...58C}. The FRB location is shown with a cross and is associated with a kpc-scale star-forming region. The low-metallicity and stellar mass of this host has suggested that FRBs are produced in the same environment (or even same kind of object) as SLSN-I and long GRBs \citep{2017ApJ...841...14M}.
  }
  \label{fig:loc}
\end{wrapfigure}

\setcounter{page}{1}
\noindent {\large \bfseries{Introduction}} The last decade has seen the rapid, concurrent development of new classes of energetic astrophysical transients. At cm-wavelength, radio telescopes have discovered 
the Fast Radio Burst\footnote{See the FRBCat for the latest total: \url{http://frbcat.org} \citep{2016PASA...33...45P}} \citep[FRB;][]{2007Sci...318..777L, 2013Sci...341...53T, 2014ApJ...790..101S},
a millisecond radio transient characterized by a dispersion measure\footnote{Dispersion measure is the line integral of the electron volume density, $DM = \int_{0}^{D} n_e dl$.} (DM) that implies an origin far outside our own Galaxy. In 2017, 10 years after the discovery of FRBs, the first FRB was conclusively associated with a host galaxy \citep[see Figure \ref{fig:loc};][]{OPT}. This association confirmed that FRBs live at distances of $\mathcal{O}(\rm{Gpc})$ and have luminosities orders of magnitude larger than pulsars and other Galactic classes of millisecond radio transients. Energetic requirements for FRBs are pushing theoretical models in new directions \citep[for a recent review, see][]{2018arXiv181005836P}.

Large optical surveys have helped characterize rare and extremely luminous classes of transients, such as the superluminous supernova \citep[SLSN;][]{2012Sci...337..927G}. SLSNe are a class of supernovae that are an order of magnitude more luminous than typical core-collapse supernovae. The ``type-I'', or hydrogen poor SLSNe, lack emission line signatures near peak luminosity, implying they are not powered by interaction with circumburst medium\footnote{The $10^{51} \, {\rm erg}$ of radiated energy of SLSNe would necessitate an enormous ejecta kinetic energy $\sim 10^{52} \, {\rm erg}$ for the interaction model to work. This large energy is likely unattainable through standard core-collapse, requiring an additional energy source even in this scenario.}. Thus some kind of ``central engine'' is required, such as accretion onto a black hole \citep[BH;][]{2013ApJ...772...30D} or the spin-down power of a young neutron star \citep[NS;][]{2010ApJ...717..245K, 2010ApJ...719L.204W}.

At high energies, gamma-ray bursts (GRBs) have long been known to require a central engine to power its relativistic jets, but the formation and nature of the engine is also not yet well defined \citep{1999ApJ...524..262M,2014MNRAS.443...67M}. ``Ultra long'' GRBs with durations exceeding thousands of seconds additionally impose severe energetic requirements \citep{2014ApJ...781...13L}. The association of a gravitational wave source, GW170817, with a short GRB and relativistic outflow was a remarkable confirmation of the fate of NS mergers \citep{1989Natur.340..126E,LIGO+17DISCOVERY, 2018Natur.561..355M}.

\textbf{Here, we consider implications of a unified framework for interpreting these transients as signatures of magnetar birth.}
This framework uses the spin-down power of a newborn magnetar \citep{2015MNRAS.454.3311M, 2016MNRAS.461.1498M} to power SLSNe, GRBs, and similar transients. At later times, magnetic activity can drive flaring and coherent radiation in the form of FRBs \citep{2013arXiv1307.4924P, 2014MNRAS.442L...9L, 2017ApJ...843L..26B, 2019arXiv190201866M}. Some predictions of this scenario have been validated by the association of ultralong GRBs \citep{2014ApJ...781...13L,2017ApJ...839...49G} with peculiar supernovae \citep{2015Natur.523..189G} or the association of a luminous radio source with an FRB \citep{2017Natur.541...58C} and a SLSN-I \citep{2019arXiv190110479E}. Furthermore, the volumetric event rate and host galaxies of SLSNe and long GRBs may be similar to that of FRBs \citep{2017ApJ...841...14M, 2017ApJ...843...84N}. As yet, there is no smoking gun of magnetar birth, but the breadth of observed phenomena suggest that \textit{we have already witnessed this process and can begin to consider its scientific impact}.

The study of young magnetars touches on the physics of NSs, BHs, and can be used to probe fundamental physics on scales ranging from the quantum to cosmological. At the smallest physical scales, the study of magnetars born in binary NS mergers has been used to characterize the equation-of-state (EOS) of dense neutron-rich matter \citep{LIGO+17DISCOVERY,2017ApJ...850L..19M,2017ApJ...850L..34B,2016PhRvD..93d4065G,2014PhRvD..89d7302L}. Forthcoming gravitational-wave (GW) observations of binary NS mergers will further revolutionize the field, but rely on {\it multi-messenger} electromagnetic (EM) followup to determine the merger fate (e.g. whether or not a magnetar was formed). Radio followup is particularly key in addressing this question, as magnetars born in NS mergers are expected to produce bright, decade-long radio emission \citep{2014MNRAS.439.3916M,2014MNRAS.437.1821M,2016ApJ...831..141F,2016ApJ...819L..22H} as well as possibly FRBs \citep{2013PASJ...65L..12T,2018PASJ...70...39Y}.

The study of radio transients will be crucial to understanding the engines of extreme transient events. The coming decade will see accelerating discoveries from funded programs for optical and radio transient discovery (e.g., ZTF, LSST, Pan-STARRS, CHIME, ASKAP, MeerKAT). Characterizing the environments in which such magnetars are born will also improve our understanding of stellar evolution and collapse, with implications to questions regarding core/envelope angular momentum coupling and magnetic field amplification in compact objects. The emission physics of FRBs may also provide clues in understanding the emission processes of pulsars, a long standing open question.

Finally, FRBs are unique probes of the intergalactic and circumgalactic medium \citep[IGM, CGM;][]{2014ApJ...780L..33M}. Radio dispersion has long been used to measure electron density in the Milky Way \citep{2002astro.ph..7156C}, but large samples of localized FRBs will extend this to intergalactic scales to probe electrons and baryons residing in the cosmic web out to high $z$ \citep{2018ApJ...852L..11S, 2019MNRAS.485..648P}. As such they may drive significant scientific progress in understanding baryon distribution and structures in galactic halos and resolving the so-called ``missing baryons problem'' at $z\lesssim2$ \citep{2014ApJ...780L..33M}. Furthermore, large samples of well-localized FRBs have been suggested as a means of addressing fundamental questions in cosmology, following suite from other astrophysical transients \citep{2018PhRvD..98j3518M,2019MNRAS.484.1637J}.

This white paper considers the scientific impact of the ``magnetar unification'' scenario. The discovery and characterization of extragalactic radio transients on timescales ranging from milliseconds to gigaseconds will offer unique opportunities and is currently the only wavelength with which FRB phenomena are associated. Below, we describe the physics of magnetars and the unification scenario. Then we highlight key observational signatures in the radio domain and the scientific progress which these may drive. Finally, we summarize the observational capabilities needed to study these signatures.

\vspace{0.3cm}
\noindent {\large \bfseries{Physics of Young Magnetars}}
Magnetars are the most magnetic objects known in the Universe. These NSs have surface magnetic field strengths of $\gtrsim 10^{14}$~G. There are roughly 29 magnetars directly observed in the Milky Way and Magellanic Clouds\footnote{See the McGill magnetar catalog at \url{http://www.physics.mcgill.ca/~pulsar/magnetar/main.html}.}. These have spin periods of several seconds and inferred ages of $\sim 10^4$~yr \citep{2017ARA&A..55..261K}. The rapid interplay of observations and theory have helped the field converge to a detailed physical model \citep{1995MNRAS.275..255T, 1998Natur.393..235K} in which the rich phenomenology of Galactic magnetars is powered by these objects' internal magnetic energy reservoir, $E_{\rm mag} \approx 3 \times 10^{49} \, {\rm erg} \left( B_\star / 10^{16} \, {\rm G} \right)^2$. For known Galactic magnetars, this magnetic energy far exceeds the magnetars' current rotational energy. Upon their initial formation however, this situation may be markedly different.


Magnetars are thought to be born in core-collapse events with high-angular momentum progenitors or in the aftermath of binary NS mergers. At birth, magnetars are thus expected have spin periods ${\rm P_0} \sim \mathcal{O}(\rm{ms})$ \citep{2015Natur.528..376M}, and a corresponding rotational energy $E_{\rm{rot}} \approx 2\times 10^{52}~\rm{erg}~(P_0/1~\rm{ms})^{-2}$, and spin-down timescale $t_{\rm sd} \approx 1.5 \times 10^4 \, {\rm s}~(B/10^{14}~\rm{G})^{-2}~(P_0/1~\rm{ms})^{2}$ \citep{2015MNRAS.454.3311M},
where $B$\ is the surface dipole magnetic field and $t$\ is the magnetar age. For $B$ $\sim$10$^{14}$-10$^{16}$ G, this luminosity can explain the required energetics and durations of SLSNe and GRBs, suggesting that millisecond magnetars are viable central-engines for these classes of transients.

The gamma-ray, optical, and radio transient properties can be used to infer magnetar properties from birth through its early years as an energetic toddler. The SLSN optical luminosity and its evolution measure the dipole magnetic field strength and birth spin period \citep{2017ApJ...850...55N}. After the ejecta becomes transparent to cm-wavelength radio emission \citep[on timescales of $\sim$decades;][]{2018MNRAS.474..573O}, the newborn magnetar may be seen as an FRB source and/or a slowly-evolving radio transient produced by its magnetar wind nebula \citep{2018ApJ...868L...4M,2019arXiv190110479E}. After $\gtrsim1$yr, this nebula would be powered by the magnetar's magnetic energy and composed of electron/ion plasma ablated from the NS crust \citep[in contrast to electron/positron rotational-powered pulsar wind nebulae;][]{2017ApJ...843L..26B}. It is even possible that the FRB emission can be used to measure the magnetar rotation period to estimate the late-time radiative power.

Binary NS mergers, which are key gravitational-wave sources, the likely progenitors of short GRBs, and an important site of $r$-process nucleosynthesis, may in certain cases form magnetar remnants as well \citep{2017ApJ...844L..19P}.
Indeed, there is evidence for such remnants in the shallow decay in the X-ray afterglow of some short GRBs \citep{2013MNRAS.430.1061R}, which may be attributed to energy injection from a newborn magnetar. Such magnetar forming mergers should also produce copious radio emission as the merger ejecta, energized by the magnetar energy input, plows into the surrounding interstellar medium \citep{2011Natur.478...82N}. Detection of this radio signature is extremely important as a calorimeter for the ejecta energy and thus remnant fate, as well as a means of probing the ambient density in the vicinity of the merger \citep{2014MNRAS.437.1821M,2016ApJ...819L..22H,2016ApJ...831..141F}. It is also possible that stable magnetars formed through this channel may eventually produce FRBs.



Local, directly observed, magnetars are relatively old ($\gtrsim 1$~kyr), so their birth properties and progenitors are not yet well understood. Furthermore, there is evidence that SLSNe, long GRBs, and FRBs are found in special environments, unlike those present in the Milky Way \citep{2006Natur.441..463F, 2015ApJ...804...90L, OPT}. This, combined with the rarity of luminous transients, suggests that magnetars born in these special environments are more ``extreme'' versions of their Galactic counterparts. If so, then the most exotic and powerful such young magnetars may only be studied as extragalactic transient sources, and such studies may in turn inform stellar evolution modelling of the progenitors and environments of such energetic compact objects.

\begin{figure}[t]
  \centering
    \includegraphics[width=0.9\columnwidth]{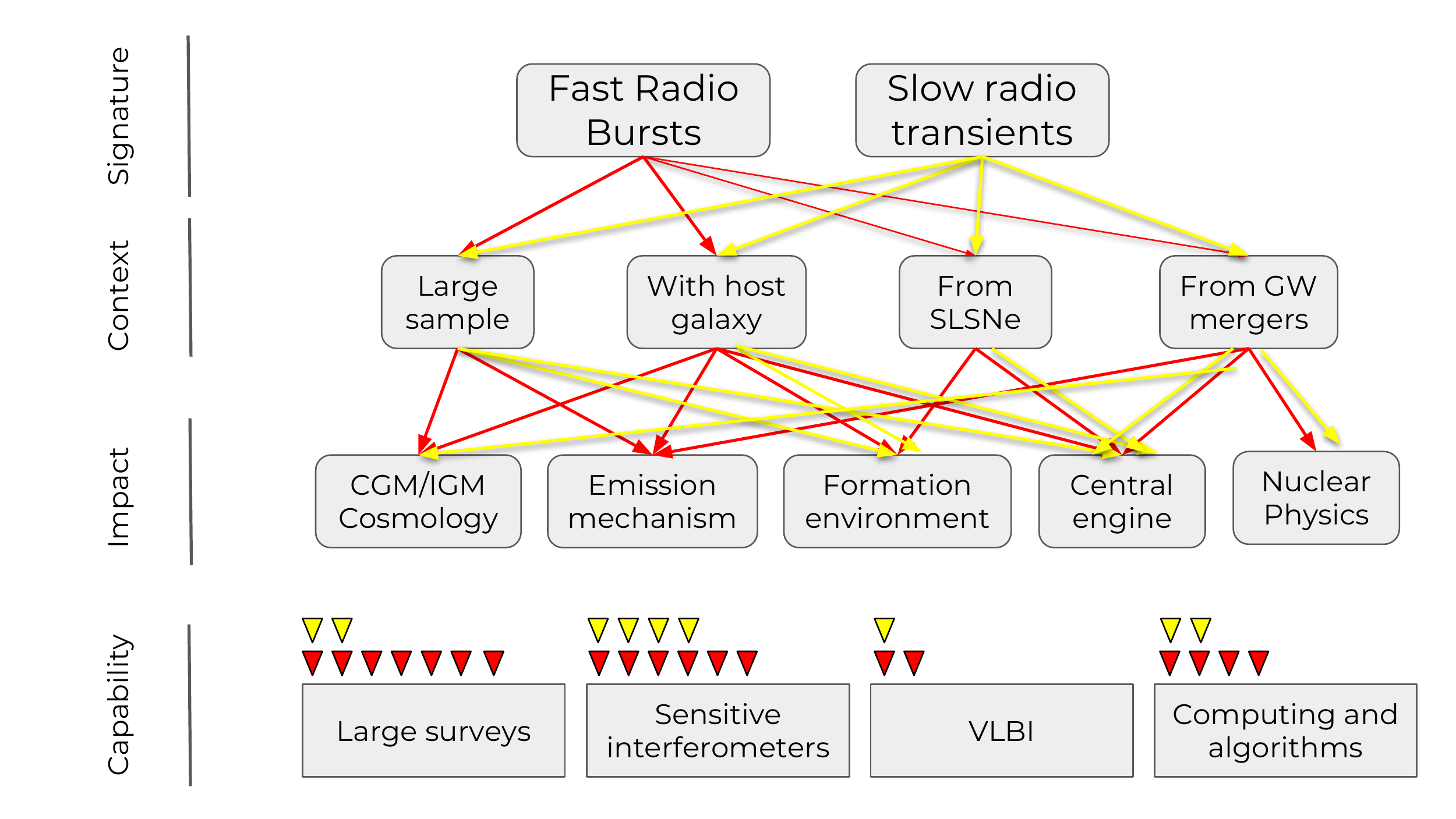}
    \caption{A diagram summarizing the association of radio transients to science topics. Each signature and context can be traced to one or more scientific impact. For each scientific impact, a triangle is added to one or more observational capability. Yellow color is associated with slow transients and red with FRBs. Some recent results have already drawn a connection between radio transients and young magnetars \citep{2015Natur.523..189G, 2017Natur.541...58C, 2019arXiv190110479E}. }
    \label{fig:diagram}
\end{figure}

\vspace{0.3cm}
\noindent {\large \bfseries{Signatures of Magnetar Birth
and their Scientific Impact}}
Figure \ref{fig:diagram} associates radio transient signatures to key science questions and the observational capabilities required to study them. Below, we describe a few, key examples drawn from this diagram.

{\bf FRB Host Galaxies:} The identification of an FRB host galaxy was critical to understanding FRBs. The measured distance to the host galaxy allowed an estimate of the baryonic content of the IGM \citep{OPT}. With $\mathcal{O}(10)$\ host galaxies, the FRB origin could be identified \citep{2017ApJ...849..162E}; $\mathcal{O}(100)$\ localizations will measure the baryon fraction of galaxy halos \citep{2019MNRAS.485..648P,2019ApJ...872...88R}. The first FRB host galaxy type suggested a connection to SLSN and long GRB \citep{2015ApJ...804...90L}, while association with a luminous, \emph{persistent} radio source has been used to infer the source age and other properties \citep{2017ApJ...843L..26B, 2017ApJ...841...14M, 2017ApJ...842...34W, 2017ApJ...839L...3K, 2018ApJ...868L...4M}. New FRB associations can test predictions for the time evolution, magnetism, and high-energy emission from magnetars \citep{2018Natur.553..182M}. Sensitive, arcsecond-resolution radio interferometers are the only proven way to find host galaxies for FRBs \citep{2017Natur.541...58C} and require novel computing architectures to solve the search problem at scale \citep{2018ApJS..236....8L,2018ApJ...868L...4M}.

{\bf Slow Radio Transients from Known SNe/GRBs:} The magnetar birth scenario predicts that magnetar-powered SNe or GRBs should produce unique $\sim$decade timescale radio transient signatures that can be used to study magnetar properties \citep{2016MNRAS.461.1498M,2016ApJ...824L..32P, 2018MNRAS.474..573O}. Dedicated follow-up has identified the first candidate radio transient associated with a SLSN-I \citep{2019arXiv190110479E}. Arcsecond-resolution radio interferometers with high sensitivity and wide frequency coverage (1--20~GHz) can distinguish this signature from foregrounds and characterize them to infer magnetar properties. Milliarcsecond-resolution radio interferometry can measure source size and surface brightness to constrain emission models \citep{EVN,2018ApJ...868L...4M}.

{\bf FRBs from Known SNe/GRBs:} Magnetar-powered SNe/GRBs may leave remnants that may also produce millisecond radio transients, FRBs. As supernova catalogs become more complete and include older supernovae, we are more likely to find FRBs from putative magnetar remnants \citep{PBS2017,2019arXiv190110479E}. Sensitive, single-dish radio telescopes are needed to search for such signatures. Innovative computing systems, such as wide-band instruments and commensal data analysis \citep[e.g.,][]{2014ApJ...790..101S}, can add transient searches for thousands of hours of observations conducted for NS mergers, SNe, and other slow transients that may also produce FRBs. 

{\bf Slow and Fast Transients from NS Mergers:} Some NS mergers are expected to leave magnetar remnants. They may spin down and collapse to a BH or they may remain as long-lived stable NSs \citep{2013MNRAS.430.1061R,2014PhRvD..89d7302L}. Magnetar spin down unleashes an enormous amount of energy into the surrounding ejecta, far exceeding the ejecta energy if a magnetar is not formed. Radio emission from the interaction of this ejecta with the surrounding ISM is an important probe of this energy, and can thus reveal the nature of the merger remnant (BH / magnetar). FRBs may be emitted from such a source at late times. In either case, follow-up with sensitive radio interferometers at a range of frequencies will infer the merger remnant fate, which is a unique probe of the uncertain EOS of dense nuclear matter \citep[e.g.][]{2017ApJ...850L..19M}.

\vspace{0.3cm}
\noindent {\large \bfseries{Capabilities for the Next Decade}}
We have shown how the magnetar unification scenario predicts a range of science applications through the study of radio transients. Here we summarize the observational capabilities needed to advance this field.

{\bf Sensitive radio interferometers:} Large radio interferometers are increasingly capable of identifying faint sources amidst foregrounds and terrestrial interference. The design of a radio telescope (e.g., antenna size, array extent, receivers) can shift its core scientific impact from one topic to another. 
A few common designs and their strengths are:
{\bf (a)} Interferometers with a large number of small dishes (e.g., Deep Synoptic Array) --- efficient survey systems for discovery and identifying large samples of FRB host galaxies;
{\bf (b)} High-sensitivity, high-frequency interferometers (e.g., ngVLA) --- powerful follow-up of luminous transients throughout local universe and ability to measure milliarcsecond sizes to characterize energetics and central engine;
{\bf (c)} Large, single-dish telescopes (e.g., GBT, Arecibo) --- versatile at characterizing FRBs (e.g., polarimetry) and enabling high-sensitivity long-baseline interferometry.
A broad investment will provide capabilities for both transient discovery and characterization, with relevance to a broad range of topics from compact object formation to multi-messenger astrophysics to galaxy evolution and cosmology.

{\bf Large Radio and Optical Surveys:}  The pioneering radio surveys of the 1990s are being updated with a new generation of surveys \citep{2015ApJ...806..224M}. The matched spatial resolution with optical surveys makes radio interferometric catalogs especially effective at identifying radio transients, as demonstrated in the discovery of FIRST J141918.9+394036 \citep{2017ApJ...846...44O,2018ApJ...866L..22L}, 
whose radio spectrum and temporal evolution are consistent with being the first orphan GRB afterglow, but the magnetar powered supernova model is also possible \citep{2015MNRAS.454.3311M}. Synoptic surveys with sensitive radio interferometers can leverage existing investments in optical transient surveys and photometric catalogs (Pan-STARRS, LSST, DESI) and large-scale spectroscopic catalogs (CLU).

{\bf Computing and algorithms:} Transient radio astronomy will be revolutionized if it can continue to develop the vision of a software-defined telescope. As computing platforms become more powerful and flexible, it will become easier to develop novel algorithms or adapt them to new uses. The CHIME telescope demonstrated this by adding FRB discovery to its science scope through dedicated computing and search software \citep{2018ApJ...863...48C,2019Natur.566..230C}. Support for software infrastructure and computing will grow capabilities in ``needle-in-the-haystack'' type of problems such as FRB discovery and the search for technosignatures. Furthermore, the richness of radio measurements -- and their association to multi-messenger counterparts -- require the implementation of sophisticated software and algorithms. Support for interdisciplinary collaborations between astronomers, statisticians, computer scientists are necessary for progress in methodology specific to astronomical data.

\clearpage

\pagenumbering{gobble}
\bibliographystyle{unsrt}

\end{document}